\documentclass[lettersize, journal]{IEEEtran}
\usepackage{graphicx} % Required for inserting images
\usepackage{amsmath,amssymb,amsfonts}
\usepackage{algorithmic}
\usepackage[nolist, withpage]{acronym}
\usepackage{pifont}
\usepackage{tabularx}
\usepackage{multirow}
\usepackage{makeidx}  % allows for indexgeneration
\makeindex
\usepackage{pifont}
\usepackage{url}
\usepackage{subfigure}
\usepackage{color}
\usepackage{multirow}
\usepackage{listings}
\usepackage[ruled,boxed, vlined,linesnumbered,resetcount]{algorithm2e}
\usepackage{tikz}
\usepackage{todonotes}
\newcommand{\MF}[1]{\textcolor{black}{#1}}
\newcommand{\FI}[1]{\textcolor{black}{#1}}

\usepackage{afterpage}
\usepackage{placeins}

\usepackage{hyperref}
\hypersetup{
      colorlinks,%
      citecolor=blue,%
      filecolor=black,%
      linkcolor=black,%
      urlcolor=black
}

\newcommand{\circled}[1]{\tikz[baseline=(myanchor.base)] \node[circle,fill=.,inner sep=1pt] (myanchor) {\color{-.}\bfseries\footnotesize #1};}

\title{2024-bier-te-fluechter}

\author{\IEEEauthorblockN{
        Moritz~Flüchter\IEEEauthorrefmark{1},
        Steffen~Lindner\IEEEauthorrefmark{1},
        Fabian~Ihle\IEEEauthorrefmark{1},
        Toerless Eckert\IEEEauthorrefmark{2},
		Michael~Menth\IEEEauthorrefmark{1}}
		
	\IEEEauthorblockA{
		\IEEEauthorrefmark{1}University~of~Tuebingen,
		Chair~of~Communication~Networks,
		72076~Tuebingen,
		Germany\\
	}
    \IEEEauthorblockA{
    \{%
        moritz.fluechter, %
        steffen.lindner, %
        fabian.ihle, %
		michael.menth\}@uni-tuebingen.de \\
    }
    \IEEEauthorblockA{
		\IEEEauthorrefmark{2}Futurewei USA,
		2560 N. First St., Suite-200 San Jose,  CA 95131
		United States of America\\
	}
    \IEEEauthorblockA{
		tte@cs.fau.de
	}
	
}

\title{Resilience for BIER-TE in Large Multicast Domains: Concept, Implementation and Performance}
\title{Extensions to BIER Tree Engineering (BIER-TE) for Large Multicast Domains and 1:1 Protection: Concept, Implementation and Performance}

\date{February 2024}

\newcommand\fig[1]{Figure~\ref{fig:#1}}

\newcommand\twofigs[2]{Figures~\ref{fig:#1} and \ref{fig:#2}}
\newcommand\sect[1]{Section~\ref{sec:#1}}
\newcommand\threesect[3]{Sections~\ref{sec:#1},~\ref{sec:#2},~\ref{sec:#3}}
\newcommand\tabl[1]{Table~\ref{tab:#1}}

\newcommand{\figeps}[3][]{%
 \begin{figure}[htb!]
  \begin{center}
   \leavevmode
      \parbox[t]{#1}{%
        \resizebox{#1}{!}{\includegraphics{figures/#2}}
      }
   \caption{#3\vspace{-0.2cm}}
   \label{fig:#2}
  \end{center}
 \end{figure}
}

\newenvironment{tab}[2]{%
 \begin{table}[htbp!]
  \begin{center}
  \caption{#2\label{tab:#1}}
}{%
  \end{center}
 \end{table}
}

\newcommand{\twofigepsDouble}[8]{
  \begin{figure*}[htb]
    \leavevmode
    \begin{center}
      \parbox[t]{#1\textwidth}{%
        \resizebox{#2\textwidth}{!}{\includegraphics{figures/#3}}
        \caption{#4}\label{fig:#3}
      }
      \hfill
      \parbox[t]{#5\textwidth}{%
        \resizebox{#6\textwidth}{!}{\includegraphics{figures/#7}}
        \caption{#8}\label{fig:#7}
     }
    \end{center}
  \end{figure*}
}

\newcommand{\twosubfigeps}[5]{
  \begin{figure}[htb]
    \leavevmode
    \begin{center}
     \subfigure[#2]{
        \label{fig:#1}
        \parbox[t]{0.9\columnwidth}{%
            \resizebox{0.9\columnwidth}{!}{\includegraphics{figures/#1}}
        }
     }
     \subfigure[#4]{
        \label{fig:#3}
        \parbox[t]{0.9\columnwidth}{%
            \resizebox{0.9\columnwidth}{!}{\includegraphics{figures/#3}}
        }
     }
    \end{center}
    \caption{#5}
  \end{figure}
}

\newcommand{\twosubfigepsFull}[5]{
  \begin{figure*}[htb]
    \leavevmode
    \begin{center}
     \subfigure[#2]{
        \label{fig:#1}
        \parbox[t]{0.47\textwidth}{%
            \resizebox{0.42\textwidth}{!}{\includegraphics{figures/#1}}
     \vspace{-1cm}
        }
     }
     \subfigure[#4]{
        \label{fig:#3}
        \parbox[t]{0.47\textwidth}{%
            \resizebox{0.42\textwidth}{!}{\includegraphics{figures/#3}}
     \vspace{-1cm}
        }
     }
    \end{center}
    \vspace{-0.5cm}
    \caption{#5}
  \end{figure*}
}

\newboolean{makevspace}
\newcommand{\cvspace}[1]{%
    \ifthenelse
        {\boolean{makevspace}}
        {\vspace{#1}}
        {}%
    }

\begin{document}

\maketitle

\begin{abstract}
Bit Index Explicit Replication (BIER) has been proposed by the IETF as a stateless multicast transport technology. 
BIER adds a BIER header containing a bitstring indicating receivers of an IP multicast (IPMC) packet within a BIER domain. 
BIER-TE extends BIER with tree engineering capabilities, i.e., the bitstring indicates both receivers as well as links over which the packet is transmitted.
As the bitstring is of limited size, e.g., 256 bits, only that number of receivers can be addressed within a BIER packet.
To scale BIER to larger networks, the receivers of a BIER domain have been assigned to subsets that can be addressed by a bitstring with a subset ID.
This approach is even compliant with fast reroute (FRR) mechanisms for BIER.

In this work we tackle the challenge of scaling BIER-TE to large networks as the subset mechanism of BIER is not sufficient for that purpose. 
A major challenge is the support of a protection mechanism in this context. 
We describe how existing networking concepts like tunneling, egress protection and BIER-TE-FRR can be combined to achieve the goal. 
Then, we implement the relevant BIER-TE components on the P4-programmable Tofino ASIC which builds upon an existing implementation for BIER.
Finally, we consider the forwarding performance of the prototype and explain how weaknesses can be improved from remedies that are well-known for BIER implementations. 
\end{abstract}
\section{Introduction}
The objective of multicast is to send at most a single packet copy over each link within a network when sending packet to multple receivers.
In IP multicast (IPMC) receivers join a multicast group whose composition is signaled to all nodes of the multicast distribution tree in order to facilitate appropriate forwarding.
This requries state maintenance in nodes in case of relevant changes.
Moreover, in the presence of a failure, lots of IPMC groups need to be recovered imposing a high signaling load on the nodes.
Therefore, the IETF has proposed Bit Index Explicit Replication (BIER) as a domain concept.
An ingress node adds a BIER header containing a bitstring indicating all receivers of an IPMC group within the BIER domain. 
BIER nodes use this bitstring to forward BIER packets along existing paths in the so-called routing underlay so that they no longer need IPMC group information. 
This makes BIER nodes stateless and more robust against failures. 
Bitstrings have limited size, e.g., 256 bits so that at most this number of receivers can be addressed by a single packet.
More receivers can be subdivided into subsets and addressed by bitstrings with subset IDs. 
On the downside, addressing receivers in different subsets requires sending different BIER packets with different bitstrings, which slightly reduces the effectiveness of multicast.
On the upside, this mechanism is compliant with BIER-Fast-Rerout (BIER-FRR), a 1:1 protection mechanism for BIER against link and node failures.

BIER-TE is a variant of BIER which also indicates links in the bitstring.
This facilitates the definition of a distribution tree in the packet header in addition to the receivers so that BIER-TE nodes can forward BIER packets along explicit paths. 
This is referred to as tree engineering (TE).
BIER-TE is very similar to BIER so BIER hardware can be utilized with only minor changes for BIER-TE forwarding. 
To protect BIER-TE against link or node failures, BIER-TE-FRR has been proposed for 1:1 protection. 
While the accommodation of receivers and links requires more bits for BIER-TE than for BIER, the definition of subsets for BIER-TE is not sufficient for scaling BIER-TE to large networks for the following reasons.
First, BIER-TE requires a connected subset of links and receivers for forwarding as BIER-TE does not leverage a routing underlay.
Second, the application of BIER-TE-FRR imposes even more constraints.

In this paper, we define how BIER-TE can be scaled to large networks including a 1:1 protection method so that resilient communication is supported for stateless multicast with tree engineering capabilities. 
The solution leverages by intention well-known networking concepts like MLPS tunneling, MPLS egress protection and BIER-TE-FRR.
MPLS may be exchanged by another forwarding technology. 
In the proposed extensions, a BIER domain is subdivided into subsets of two-connected links and nodes. 
When an ingress node receives an IPMC packet with receivers in a specific subset, the IPMC packet is equiped with a BIER header for that subset and tunneled to an ingress node of the subset. 
From there, the BIER packet is forwarded with native BIER-TE. 
Within subsets, normal BIER-TE-FRR protects against failures.
Challenges arise if the subset ingress fails. 
We leverage egress protection to divert the BIER packet from a point of local repair (PLR) to a backup subset ingress which applies a variant of BIER-TE-FRR to handle the packet.
Thus, we define the novel node functionalities: domain ingress, subset ingress, PLR, and backup subset ingress while forwarding nodes in subsets just apply normal BIER-TE forwarding.
To show the feasibility of the proposal, we implemented these functions on the Intel Tofino using the P4 programming language and show that they support 100 Gb/s per port.
Only replication nodes have a lower throughput, which is in line with existing BIER implementations and we explain how existing speedup methods for BIER are also applicable to BIER-TE. 
While the selection of subsets for BIER has been well understood, the selection of subsets for BIER-TE remains an open problem.
We describe properties for such subsets and the need for additional virtual links such that BIER-TE can provide the same level of resilience as BIER.
These findings serve as input for future work designing selection algorithms for BIER-TE subsets. 

The paper is structured as follows.
We first provide a short introduction to \ac{BIER} and \ac{BIER-TE} in \sect{bier_intro}.
Then, we summarize advances in scalable multicast and resilience mechanisms for \ac{BIER} and \ac{BIER-TE} in \sect{related_work}, 
Then, we introduce the fundamentals of \ac{BIER} for 
In \sect{p4}, we give a brief introduction to the P4 programming language and the Intel Tofino\texttrademark{} switching ASIC.
In \sect{scalability}, we elaborate in detail on \ac{BIER-TE} subset tunneling.
We then provide an overview of the P4 prototype implementation in \sect{implementation}.
In \sect{evaluation} we evaluate the IPMC throughput of our P4 prototype.
Last, we summarize the contributions of this work in \sect{conclusion}.

%\todo[inline]{Add section labels}
%\todo[inline]{We could calculate the scalability issues for all different domain types}

%\todo[inline]{SL: Noch recht unverständlich was das Problem ist und was das Ziel des Papers ist.}

\iffalse
ac{BIER-TE}~\cite{TODO} was introduced by the IETF as a stateless transport mechanism for \ac{IPMC} traffic that supports traffic engineering.
\ac{BIER-TE} leverages domains that do not require dynamic signaling.
Domain ingress nodes push a \ac{BIER-TE} header to any incoming \ac{IPMC} packet containing the path through the domain and egress nodes.
\ac{BIER-TE} forwarding nodes use this header to determine the next hops for a packet.
When a packet reaches the edge of the domain, the egress node removes the \ac{BIER-TE} header, and the standard \ac{IPMC} forwarding is applied.
This approach reduces the signaling load on forwarding nodes since only ingress and egress nodes handle \ac{IPMC} signaling.

Encoding the multicast path in the \ac{BIER-TE} header causes a scalability issue in larger networks.
The more nodes in a \ac{BIER-TE} domain, the larger the header and overhead becomes.
Splitting the domain into smaller subsets is possible via a \ac{SI} but selecting these subsets is a complex problem.
The underlying cause is that the subset has to include the path from the domain ingress node.
Adding a single path from the ingress does not allow any \ac{TE} to the subset.
Adding multiple paths re-introduces a scalability issue.
\fi
\section{Preliminaries on BIER and BIER-TE}\label{sec:bier_intro}
\MF{
In this section, we give a short introduction to \ac{BIER} and \ac{BIER-TE}.
We first explain the concept of BIER and how packets are forwarded through a BIER domain.
Then, we elaborate on how BIER subsets improve scalability and BIER \acf{FRR} improves resilience.
Last, we explain the origin and goal of \ac{BIER-TE}.
}

\MF{
\subsection{\acf{BIER}}
\ac{BIER}~\cite{rfc8279} is a protocol developed to efficiently forward \acf{IPMC} traffic.
With regular \acf{IPMC}, forwarding nodes need to store the next hops for every multicast group.
Changes to a multicast group need to be propagated through the entire network.
This introduces a significant signalling overhead in networks with frequent changes.
In contrast, \ac{BIER} introduces stateless domains where only the domain ingress has to maintain an updated information base.
Therefore, it can support many multicast groups even if they have a high churn rate.
}

\subsubsection{Domain Architecture}
\MF{
The layered architecture of a \ac{BIER} network is visualized in \fig{bier_layer.pdf}.
\ac{IPMC} sources and destination are located in the upper layer.
Below that is the \ac{BIER} layer which handles packet forwarding and replication.
All nodes taking part in the \ac{BIER} forwarding are located within the same \ac{BIER} domain.
The paths for the forwarding stem from the routing underlay.
}

\figeps[\columnwidth]{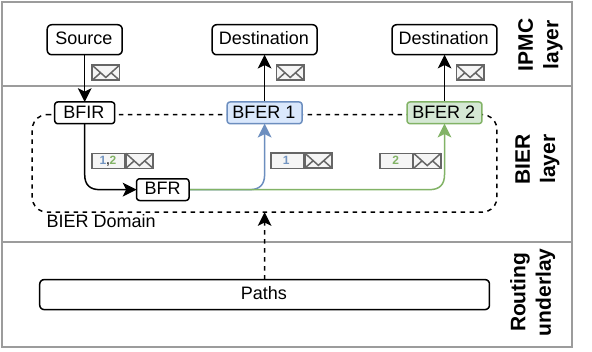}{The layered \ac{BIER} architecture as presented by~\cite{MeLi2020}. Traffic sources and destinations are located in the \ac{IPMC} layer. In the BIER layer, BIER nodes forward and replicate the \ac{IPMC} packets based on paths from the routing underlay.}

\MF{
Each \ac{BIER} domain has three types of forwarding nodes.
\acp{BFIR} are located at the entry points of the \ac{BIER} domain.
They receive \ac{IPMC} packets, add a \ac{BIER} header, and forward packets into the domain.
Their counterparts are the \acp{BFER} which remove the \ac{BIER} header and forward the packet back to the \ac{IPMC} layer.
All \ac{BIER} nodes in between the \acp{BFIR} and \acp{BFER} are called \acp{BFR}.
They replicate and forward packets through the domain based on the contents of the \ac{BIER} header.
A \ac{BIER} node may also act as a combination of two or more node types, e.g., as a \ac{BFIR} and \ac{BFR}.
}

\subsubsection{Forwarding logic}
\MF{
This forwarding through a \ac{BIER} domain is based on a bit vector contained in each \ac{BIER} header.
Each bit in the so-called \acf{BS} is assigned to a \ac{BFER} in the domain.
If the bit is set then a copy of the packet needs to be delivered to that \ac{BFER}.
The forwarding information in each \ac{BFR} is stored in the \acf{BIFT}.
It contains the reachable \acp{BFER}, i.e. bits in the \ac{BS}, per next hop.
}

%When a \ac{BFR} receives a packet, it matches the active bits against the so-called \ac{BIFT}.
%The \ac{BIFT} contains the reachable \acp{BFER}, i.e. bits in the \ac{BS}, per next hop.
%If at least one active bit overlaps, the packet is replicated and forwarded to the corresponding next hop.
%Further, it unsets the bits of all \acp{BFER} not reachable over the next hop to prevent loops.
%These steps are repeated until all active bits have been processed.
\subsection{\ac{BIER} Subsets}
\MF{
The \ac{BIER} \ac{BS} contains a bit for \ac{BFER} in the domain and does not scale well in large networks.
To address this, the \ac{BIER} domain can be subdivided into subsets.
The subsets themselves can be selected arbitrarily and do not have any topological constraints.
The \ac{BIER} header contains a \acf{SID} field that uniquely identifies the destination subdomain.
Therefore, the \ac{BIER} \ac{BS} only addresses \acp{BFER} in a single subset instead of all \acp{BFER} in the domain.
}

\MF{
When a \ac{BFIR} receives an \ac{IPMC} packet, it determines the subset of the destination \acp{BFER}.
If the \acp{BFER} are in different subsets, the \ac{BFIR} creates separate packets for each subset.
\acp{BFR} forwarding the packets match on the combination of \ac{SI} and \ac{BS} to determine the next hops.
To that end, the \ac{BIFT} contains separate entries for each configured subset in a subdomain.
}

\subsection{\ac{BIER} \acf{FRR}}
\MF{
\ac{BIER} FRR~\cite{ietf-bier-frr-04} protects against the failure of links and nodes in the \ac{BIER} domain.
When a \ac{BFR} detects a failure it becomes the \acf{PLR}.
If a link fails, \acf{PLR} encapsulates the packet in an IP header and tunnels the packet to the next hop.
The next hop decapsulates the packet and resumes regular BIER forwarding.
If a node fails, the \acf{PLR} has to deliver the packet to each next hop of the failed node, the so-called \acf{NNH}.
To that end, it determines which \acp{NNH} are part of the distribution tree and needs to receive a copy of the packet.
The \ac{PLR} then tunnels the packet to each of the \ac{NNH} via IP encapsulation.
The RFC~\cite{ietf-bier-frr-04} defines an extended BIFT that holds the required FRR information.
}

\subsection{From BIER to BIER-TE}
\MF{
\ac{BIER-TE}~\cite{rfc9262} is derived from the concept of \ac{BIER} and adds support for traffic engineering.
It is designed to run efficiently on the same hardware as \ac{BIER} and uses the same header format.
However, the \ac{BIER-TE} header contains the entire multicast distribution tree for a packet.
To that end, all links and \acp{BFER} are assigned a bit in the \ac{BIER-TE} \ac{BS}.
If a link bit is set, then the packet has to traverse the corresponding link.
\ac{BIER-TE} \acp{BFR} send a packet copy over each adjacent link whose bit is set in the \ac{BS}.
}

\section{Related Work}\label{sec:related_work}
We first summarize existing resilience mechanisms for both \ac{BIER} and \ac{BIER-TE}.
Then, we give a short overview of related work on scalable multicast solutions.

\subsection{Resilience Mechanisms for \ac{BIER} and \ac{BIER-TE}}
Braunt et al.~\cite{BrHa20172} examine resilience mechanisms for \ac{BIER} and \ac{BIER-TE}.
They present a 1+1 protection feature based on \acp{MRT}.
Their proposed \ac{MoFRR} for \ac{BIER} doubles the traffic load but requires less additional capacity compared to \ac{BIER} and \ac{BIER-TE} \ac{FRR}.
Further, they compare three different \ac{FRR} approaches and recommend \ac{BIER}-in-\ac{BIER} encapsulation as the preferred method.
Similarly, Merling et al.~\cite{frr_comp} compare \ac{LFA} and tunnel-based \ac{BIER} \ac{FRR}.
They find shortcomings of \ac{LFA}-based \ac{FRR} and proposed extensions to the standard.
However, both works do not include an implementation and evaluation of the concepts.

In a follow-up work, Merling et al.~\cite{MeLi2020} propose a version of \ac{BIER} \ac{FRR} that leverages the \ac{FRR} mechanism of the routing underlay.
The proposed approach supports both link and node protection.
They evaluate the \ac{FRR} mechanism via a P4 implementation on the BMv2 software switch.
In a virtual testbed, their \ac{FRR} implementation greatly reduces the restoration time compared to the convergence time of the routing layer.
Further, Merling et al.~\cite{MeLi2021} publish a hardware evaluation of \ac{BIER} \ac{FRR} on the Intel Tofino\textsuperscript{TM}.
They confirm the reduced restoration time when using \ac{FRR}.
However, their implementation throughput is limited by the number of next hops and available physical recirculating ports.
The work of Merling et al. has contributed directly to the current internet draft of \ac{BIER} \ac{FRR}~\cite{bierfrr}.

The limited throughput is addressed by Lindner et al.~\cite{LiMe2023}.
They improve it by clustering ports in multicast groups and creating multiple packet copies in a single iteration.
Therefore, their approach drastically reduces the total number of recirculations.

There are currently two internet drafts for \ac{BIER-TE} \ac{FRR} by Eckert et al.~\cite{bier-te-draft1} and Chen at al.~\cite{chen-bier-te-frr-07}.
These approaches are similar for link protection but differ greatly for node protection.
Eckert et al.~\cite{bier-te-draft1} propose an approach in which packets are distributed to the \ac{NNH} using backup \ac{BIER-TE} distribution trees.
These trees are either applied to the original \ac{BS} or used to encapsulate the packets in an outer \ac{BIER-TE} header.
Chen at al.~\cite{chen-bier-te-frr-07} argue that the construction of these backup multicast trees is complex and may lead to duplicate packets.
With their approach, the original packet is copied for each \ac{NNH} and each copy is tunneled to a \ac{NNH} over \ac{BIER-TE} encapsulation.
However, this causes an increased traffic load since the \ac{FRR} traffic uses unicast and not multicast.
Therefore, we focus on the internet draft of Eckert et al. for this work.

\subsection{Scalable Multicast}
Scalable and efficient multicast has been an active research topic with many contributions.
For this work, we focus on research on BIER and header encodings similar to BIER.

\subsubsection{BIER research}
Merling et al.~\cite{MeSt23} analyze the difference in traffic volume between \ac{IPMC}, \ac{BIER}, and unicast in different network topologies.
They find that \ac{BIER} produces more traffic than \ac{IPMC} but still greatly improves on regular unicast.
Further, they investigate the scalability of \ac{BIER} subsets and a suitable subset selection algorithm.
They conclude that subsets increase \acp{BIER} scalability but must be chosen carefully.

Lie et al.~\cite{hawkeye} leverage deep reinforcement learning to build multicast trees for \ac{BIER-TE}.
Their proposal also considers information about recorded streams in the network.
Eckert et al.~\cite{eckert-bier-rbs-00} find that \acf{BIER-TE} does not scale well for large networks.
Therefore, they propose \ac{RBS} which re-engineers the \ac{BIER-TE} header to improve performance and scalability.
Similarly, Diab et al.~\cite{yeti} propose YETI which outperforms \ac{BIER-TE} concerning header size, scalability, and performance.
However, both \ac{RBS} and YETI require multiple packet cycles per replication.

\subsubsection{Encodings similar to BIER}
Li et al.~\cite{li2013scaling} adjust \ac{IPMC} for data center topologies.
Their approach leverages a centralized controller that partitions the network and locally aggregates multicast groups.
In an exemplary setting with 27.648 servers, their \ac{IPMC} supports 100k multicast groups.
Elmo~\cite{shahbaz2019elmo}, Bert~\cite{alqahtani2020bert}, and Ernie~\cite{alqahtani2021ernie} also focus on multicast efficiency in data centers.
These approaches encode multicast data in the packet header to greatly increase the number of supported multicast groups and reduce traffic overhead.
However, they rely on the special data center topology and do not consider the impact of frequent changes to multicast groups.

\par\medskip

In contrast to existing improvements for \ac{BIER-TE} like \ac{RBS}~\cite{eckert-bier-rbs-00} and YETI~\cite{yeti}, we propose an approach that does not change the header encoding.
Therefore it can be implemented efficiently and without additional recirculations similar to the works of Merling et al.~\cite{MeLi2021} and Lindner et al.~\cite{LiMe2023}
Other encodings that use a different header format than BIER are limited to special topologies or do not consider high churn rates in multicast groups.
Therefore, they do not have the same benefits as \ac{BIER} and \ac{BIER-TE}.
\section{\ac{BIER-TE}}\label{sec:bierte}
In this section, we introduce \ac{BIER-TE} which extends the \ac{BIER} protocol with \acf{TE} capabilities.
First, we explain the concept of \acs{BIER} and its advantages over standard \ac{IPMC}.
Then, we introduce \acf{BIER-TE}, its forwarding procedure, and \ac{BIER-TE} subsets.
Last, we explain the resilience mechanism \acf{FRR} for \ac{BIER-TE}.

\subsection{Forwarding Example}\label{sec:traffic_engineering}
%\ac{TE} describes the capability to actively define the routes that traffic flows take through a network.
\acf{BIER-TE}~\cite{rfc9262} leverages the same header format as regular \ac{BIER}.
However, bits in a \ac{BIER-TE} header can address either a \ac{BFER} or a link in the domain.
If the bit of a link is set in a \ac{BS}, the packet must traverse the corresponding link.
\fig{bierte_forwarding.pdf} illustrates the forwarding based on the header in a BIER-TE domain.

\figeps[\columnwidth]{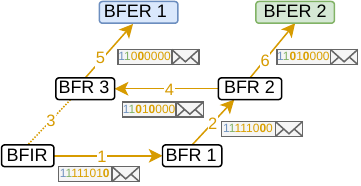}{A packet forwarding example in a \ac{BIER-TE} domain. The link numbers represent their bit position (starting from the least significant) in the header. For readability, the most significant bits belong to the \acp{BFER}.}

The packet originates from the \ac{BFIR} and is destined towards both \acp{BFER}.
For \ac{BFER} 1, the configured path is over links $\{1,2,4,5\}$ and for \ac{BFER} 2 over $\{1,2,6\}$.
To this end, the bits for all mentioned links and \acp{BFER} are set in the initial header added to the packet.
Since link 3 should not be traversed, the corresponding bit is never set.

%\todo[inline]{Nummerierung hinzufügen um BS state im Beispiel zu referenzieren}
When a BFR receives a packet it checks which bits for its adjacent links are active.
For each active bit, the packet is replicated and forwarded over that link.
The bit for the link itself is unset before forwarding to prevent the packet from traversing the link again.
For example, \ac{BFR} 1 unsets the bit for link 2 before transmitting the packet.
In the example, \ac{BFR} 2 sends a copy of the packet over links 4 and 6.

\subsection{Forwarding Logic}
%\todo[inline]{Exlain SI bit in header and meaning}
The forwarding of a \ac{BIER-TE} packet uses a single table, the \ac{BIFT}.
Each entry in the \ac{BIFT} corresponds to one of the \acp{BFR} connected links.
Hence, the entries match on set link bits in the \ac{BIER-TE} \ac{BS}.
\acp{BFER} have an additional entry that matches their assigned bit.
Each \ac{BIFT} entry also contains a \ac{F-BM} and forwarding action.
An exemplary \ac{BIFT} for \ac{BFR} 2 in \fig{bierte_forwarding.pdf} is shown in table \tabl{bift}.

\begin{tab}{bift}{An example \ac{BIFT} referencing \ac{BFR} 2 from \fig{bierte_forwarding.pdf}}
\small
\begin{tabularx}{\columnwidth}{lll}\hline
Key      & F-BM     & Forward Action            \\ \hline
00000010 & 11101010 & forward\_connected(BFR 1)  \\
00001000 & 11101010 & forward\_connected(BFR 3)  \\
00100000 & 11101010 & forward\_connected(BFER 2) \\ \hline
\end{tabularx}
\end{tab}

%Same structure as BIER BIFT, usable on same hardware
%However, logic is simpler, packet is fowarded solely based on set link adjacency
%Thus FBM is only used to unset the corresponding link

The \ac{F-BM} is used to clear the adjacencies of the current \ac{BFR} from a packet to prevent loops.
Applying it via a bitwise AND to the \ac{BS} unsets all bits of the multicast tree that have already been considered.
The forwarding action field determines where the outgoing packet is sent.
With connected forwarding, the packet is sent directly to the next hop over Layer-2.
Routed forwarding is used if there are non-\ac{BIER-TE} intermediate hops and uses Layer-3.
Load balancing is possible via \ac{ECMP} forwarding where the next hop is chosen based on the output of a hash function.
Last, a decap action is used in entries for \ac{BFER} bits to signal the decapsulation and passing to the \ac{IPMC} layer.

The processing of a packet using the \ac{BIFT} is defined in RFC 9262~\cite{rfc9262} as follows.
The \ac{BFR} iterates over all set bits in the \ac{BIER-TE} \ac{BS}.
If there is an entry in the \ac{BIFT} for that bit, the packet is replicated for forwarding over that link.
The corresponding \ac{F-BM} is applied to the \ac{BS} in the copied packet via a bit-wise AND operation.
Further, the matched link bit is unset in the \ac{BS} as mentioned in \sect{traffic_engineering}.
Then, the packet is forwarded as defined by the forwarding action.

\subsection{BIER-TE FRR}
%\todo[inline]{Change structure: A. BIER-TE FRR ide B. FRR by Toerless C. FRR by Chen. Add Chen vs Toerless evaluation}
%We first explain the basis of FRR
%Then, we elaborate on two drafts for BIER-TE FRR
%First by Toerless et al, then by Chen et al.

%\subsubsection{\ac{FRR} in Multicast}
The \ac{BIER-TE} \ac{FRR} mechanism proposed by Eckert et al.~\cite{bier-te-draft1} can protect against link and node failures.
When a \ac{BFR} detects the failure of an adjacency it becomes the \ac{PLR}.
The \ac{PLR} compensates for the failure by rerouting the packet around the failure.
This rerouting process depends on whether the \ac{PLR} protects against link or node failure as visualized in \twofigs{link_protection.pdf}{node_protection.pdf}.

\twosubfigepsFull
{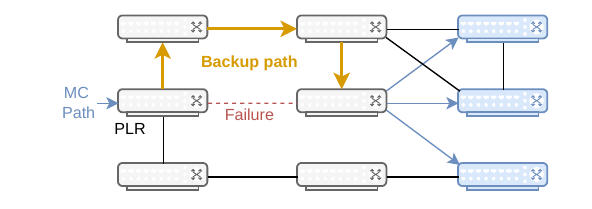}{Link protection.}
{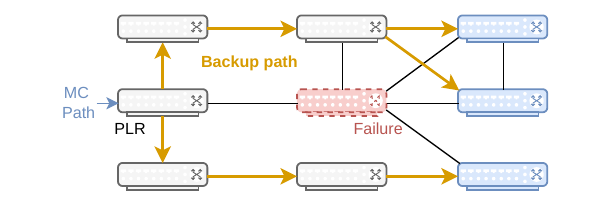}{Node protection.}
{Examples for node and link protection with \ac{BIER-TE} FRR. With link protection, the packet is rerouted to the next hop via an alternative path. With node protection, the packet is delivered to all relevant next-next hops of the failed node.}

\subsubsection{Link Protection}
\fig{link_protection.pdf} illustrates the handling of a link failure.
The \ac{PLR} assumes only the link failed and the next hop \ac{BFR} is fully operational.
Therefore, rerouting the packet to the next hop via an alternate disjoint path is enough to restore an operational path.
To that end, the backup path is applied as a tunnel, e.g. via \ac{BIER}-in-\ac{BIER}, or directly encoded into the original \ac{BS}.

\subsubsection{Node Protection}
The node protection functionality is illustrated in \fig{node_protection.pdf}.
When the \ac{PLR} detects a failure, it assumes that the next hop \ac{BFR} is not operational.
Therefore, the packet has to be forwarded to all next hops of the failed node, the so-called next-next hops.
As with link protection, the \ac{PLR} can use tunneling or rewrite the \ac{BS} to apply the backup routes.

\par\medskip

The information required for \ac{FRR} is stored in the so-called \ac{BTAFT}.
It matches on the failed adjacency in the \ac{BS} and stores the so-called \textit{ResetBitmask} and \textit{AddBitmask}.
Both have the same data type and length as the \ac{BS}.
The \textit{ResetBitmask} contains the bits that should be deactivated in the header.
E.g., in the case of link protection, the bit for the protected link is active in the \textit{ResetBitmask}.
The \textit{AddBitmask} stores the encoded backup path.
Once these masks are applied to the packet, the standard \ac{BIER-TE} forwarding continues.
\section{P4 and the Intel Tofino™ Switching ASIC}\label{sec:p4}
\FI{We briefly introduce} the P4 programming language and the Intel Tofino™, a P4-programmable switching ASIC.
We first describe concepts of the P4 language relevant to this work and then outline the architecture of the Intel Tofino™ switching ASIC.
\subsection{The P4 Programming Language}
With the \ac{P4} language~\cite{p4}, network operators can describe the data plane of P4-programmable devices, so-called targets.
A target implements a specific architecture which may be software-based or hardware-based.
The functionality of architectures can be extended by externs, such as registers and counters.
In \acs{P4}, a target-specific compiler maps the \acs{P4} program to the packet processing pipeline of a \acs{P4}-programmable device.

The architecture of a target defines intrinsic metadata that is added per packet.
Intrinsic metadata contains information such as the ingress and egress port, and timestamps.
Further, user-defined metadata can be leveraged to store values during the processing of a packet.
Metadata is only valid during the pipeline processing and does not leave the pipeline of the switch.

The pipeline of a \acs{P4} switch architecture contains a programmable packet parser, several control blocks with \acp{MAT}, and a packet deparser.
\FI{The architecture of the Intel Tofino™ switching ASIC is called \ac{TNA}~\cite{TNA}.
The pipeline of the \acs{TNA} contains an ingress control block and an egress control block, each with its own parser and deparser.
\fig{tna} illustrates the pipeline of the \acs{TNA}.}

\figeps[0.5\textwidth]{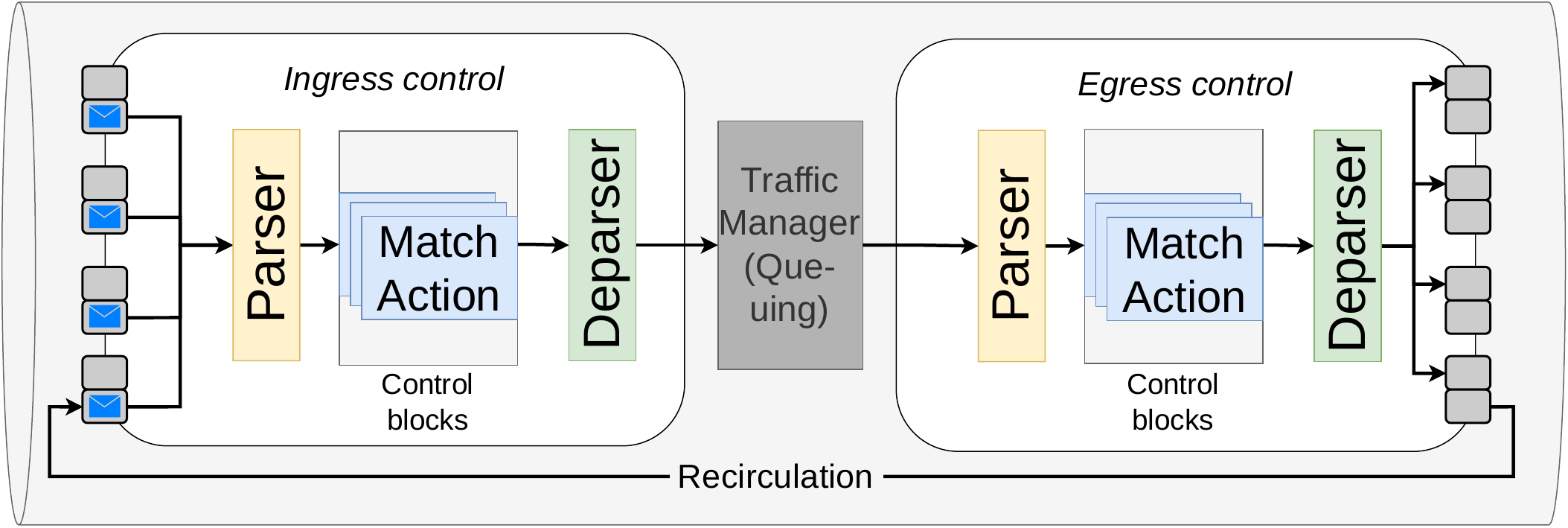}{The \acs{TNA} consists of an ingress block and an egress block separated by the traffic manager. Each block has its own programmable packet parser, \acp{MAT} and deparser.}

The packet parser extracts header information from packets according to user-defined states and transitions modeled in a finite state machine.
\FI{The control blocks access and manipulate header information extracted in the packet parser to make forwarding decisions.}
The remaining payload of a packet, i.e., the part not parsed by the parser, remains untouched and is inaccessible in the control blocks.

The control blocks of a \acs{P4} program define the packet processing operations that are applied to a packet. 
In a control block, branching constructs, simple logical and arithmetic expressions, and \acfp{MAT} can be used to describe an algorithm.
%\acp{MAT} describe the packet processing algorithm in control blocks.
The basic principle of a \acs{MAT} is illustrated in \fig{mat} and briefly described below.

\figeps[0.4\textwidth]{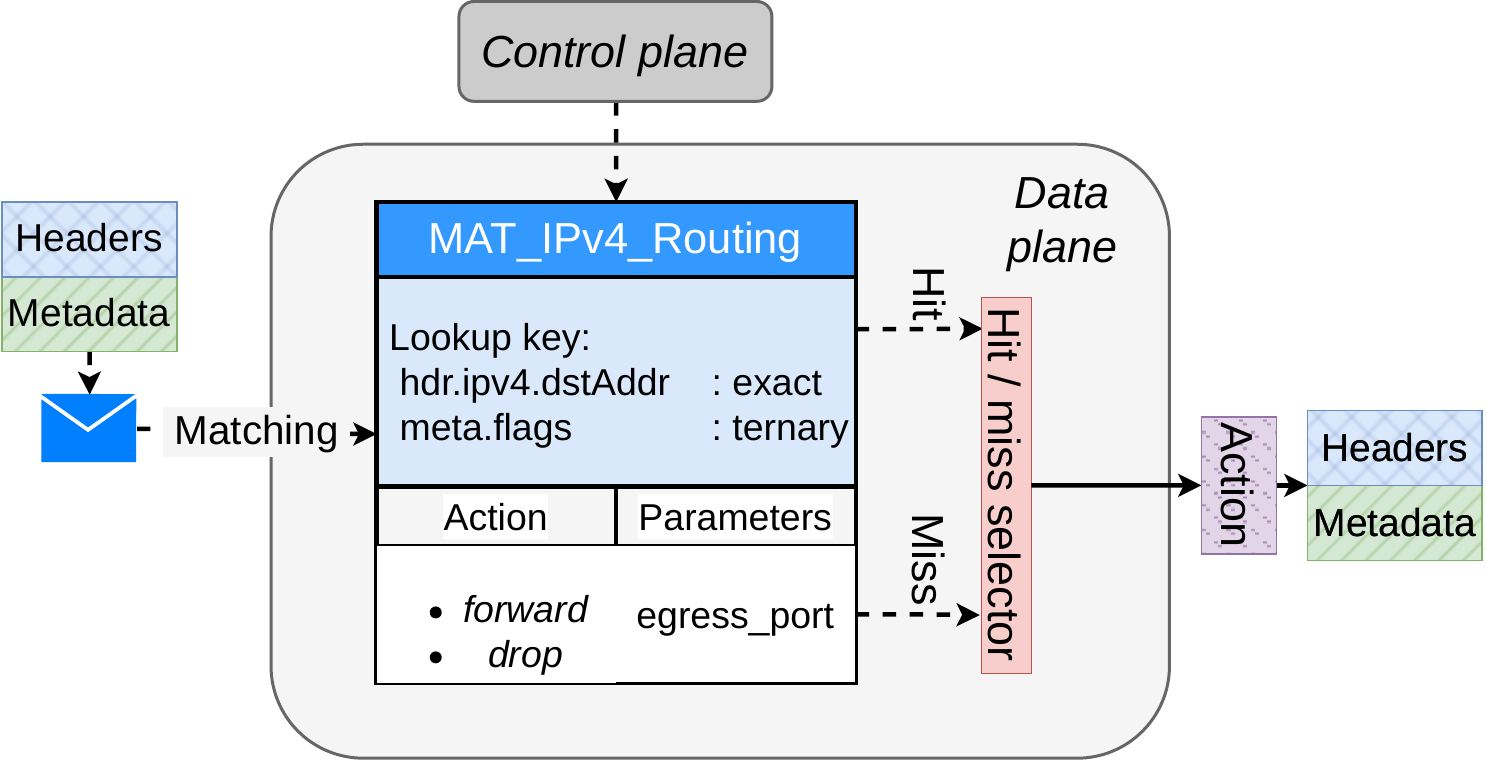}{A \acs{MAT} consists of a composite lookup key that is applied to a packet. If the lookup key matches, a predefined action is executed. The scheme of the \acs{MAT} is defined in the data plane while its content is filled by the control plane.}

A \acs{MAT} contains a lookup key composed of header and metadata fields that is matched against each packet. 
In P4, there are different matching types of lookup keys, e.g., exact and ternary.
The ternary match provides a value and a mask.
A ternary entry matches if the bitwise and operation with the field of the packet and the mask results in the specified value in the table.
On a match, a predefined action is performed in the data plane.
Such an action can manipulate the header and metadata fields of a packet to make a forwarding decision.
While the data plane defines the schema of \acp{MAT} and their corresponding actions, the control plane fills in the entries of a \acs{MAT}.

Loops do not exist in a \acs{P4} program to facilitate line-rate processing. 
Recirculation is a concept that allows to model loop behavior \FI{over multiple packet cycles.}
In a recirculation, a packet is modified while traversing the pipeline and is then sent back to an ingress port.
Then, the packet is processed again in the pipeline.
\FI{Sending a packet through the pipeline multiple times requires bandwidth capacity and adds a small, constant, known time to the processing delay.}
The capacity for a recirculation has to be provisioned.
To that end, \FI{internal} recirculation ports exist in hardware targets such as the Intel Tofino™.
If more ports are required for recirculation than internal recirculation ports are available, normal ingress ports can be configured in loopback mode to be leveraged for recirculation.

After processing a packet, the relevant headers are emitted in a user-defined order by the packet deparser.

\FI{In addition, the \acs{TNA} provides an internal packet generator.}
The internal packet generator can be configured by the control plane to \FI{generate} packets.
\FI{The generation of} packets can be triggered by various events, such as a periodic timer, a recirculation, or a port-down event.
These packets are sent out through the internal traffic generation ports.
%\todo[]{Mehr Infos zu Pipe und APP ID falls mehr gebraucht wird}

Static multicast groups in P4 facilitate packet replication to multiple egress ports. 
The control plane configures a multicast group consisting of a multicast group identifier and a set of egress ports.
The data plane assigns a multicast group identifier, e.g., by using a \acs{MAT}, to a packet in the ingress control block.
The packet is then replicated after ingress processing to the egress ports associated with the assigned multicast group identifier.
This is a local mechanism in a forwarding device to forward a packet to multiple egress ports and should not be confused with IPMC.
For this reason, we will refer to this mechanism as \textit{static multicast groups} in the following.

%Example: https://atlas.cs.uni-tuebingen.de/~menth/papers/Menth22-Sub-2.pdf Section IV
\section{BIER-TE Scalability}\label{sec:scalability}
We first elaborate on the scalability problem posed by \ac{BIER-TE} with \acp{SI}.
Then, we introduce the concept of subdomain tunneling to address this issue.
Further, we elaborate on the realization of the concept and the required ingress protection mechanism.
Last, we provide requirements and optimization goals for future work on an algorithm to select subsets.

\twofigepsDouble
{0.49}{0.5}{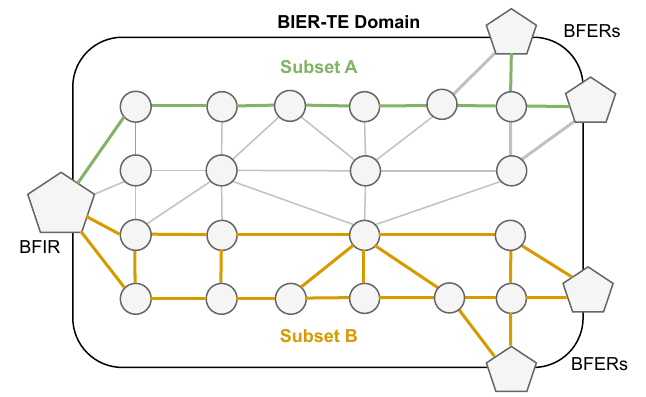}{Usage of SI in BIER-TE example, shows how links to subset have to be part of subset as well. The upper subset uses only one path and does not allow any TE on the way to the subset. 
The lower subset allows TE but the header size is increased.}
{0.49}{0.5}{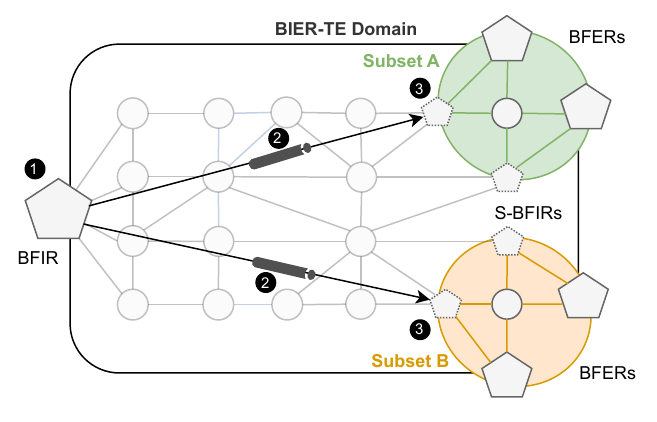}{Subset tunneling concept as presented in this work. For simplicity only subset A and B have BFERs.
Here, the \ac{BFIR} tunnels the packets to each \ac{S-BFIR} which then forwards the packets into their subset.}

\subsection{Problem Statement}\label{sec:problem_statement}
Using a \ac{SI} in combination with \ac{BIER-TE} reduces the required number of bits in the \ac{BS}.
However, this approach does not scale as effectively with \ac{BIER-TE} as it does with regular \ac{BIER}.
In BIER, all recipients encoded in a BIER packet need to be part of the same SI. 
In contrast, BIER-TE requires that all involved links, i.e., the multicast distribution tree, are encoded in the BIER-TE packet.
Consequently, all involved links must be part of the same SI.
This introduces a trade-off between the size of the distribution tree and the \ac{BS} size.
\fig{bierte_subset_scalability.pdf} illustrates this issue.

The \ac{BIER-TE} domain consists of one \ac{BFIR}, four \acp{BFER}, and twenty \acp{BFR}.
For simplicity, we only consider two subsets that do not overlap.
Subset A comprises two \acp{BFER} that are connected to the \ac{BFIR} over eight links.
The subset is only one-connected, i.e., there is only one possible path from \ac{BFIR} to \ac{BFER}.
In contrast, Subset B is two-connected which enables \ac{TE} and \ac{FRR}.
It contains two \acp{BFER} and twenty links.
For subset A the required \ac{BS} length is 10 bits and subset B requires 22 bits.
%We further analyse these scalability issues using the topologies presented in \cite{MeSt23}
%Table X contains domain topologies, upper limit for number of BFERs in a subdomains

\subsection{Concept}\label{sec:concept}
We propose an alternative distribution mechanism to enhance the scalability of \ac{BIER-TE}.
In this mechanism, each subset is assigned one or more subset ingress routers or \ac{S-BFIR}.
Packets from the \ac{BFIR} are tunneled to the \ac{S-BFIR} of each destination subset.
The \ac{S-BFIR} removes the tunneling header and forwards the packet within the subset using standard \ac{BIER-TE} and \acp{SI}.
We assume that \acp{S-BFIR} are no transit nodes, i.e., they only serve as ingress node, not as a \ac{BFR} for traffic from other \acp{S-BFIR}.
\fig{scalability_concept.pdf} illustrates this concept.

First, the \ac{BFIR} receives the \ac{IPMC} packet, determines the \acp{BFER}, and identifies the corresponding subsets \circled{1}.
It then creates a packet for each subset.
The \ac{BIER-TE} header of each packet contains only the links and \acp{BFER} in the destination subset.
An outer tunneling header is added, and the packet is tunneled to the \ac{BFIR} \circled{2}.
Upon reaching the \ac{S-BFIR}, the encapsulation is removed, and normal \ac{BIER-TE} forwarding is applied \circled{3}.

\subsection{Subset Tunneling}\label{sec:subset_tunneling}
There are multiple possibilities for a P2P tunneling protocol used with \ac{BIER-TE} subsets.
P2MP is not required since each subset receives a packet copy with a different \ac{BIER-TE} header.
We present three exemplary tunneling approaches: IP, \ac{BIER-TE}-in-\ac{BIER-TE}, and \ac{SR}/\ac{MPLS}.

\subsubsection{IP}
With IP tunneling, the packet is encapsulated in a Layer-3 header and forwarded by the routing underlay.
The destination address of the Layer-3 header is that of the \ac{S-BFIR}.
Therefore, the \acp{S-BFIR} of each subset need valid Layer-3 addresses.
While this approach is simple, it does not support any \ac{TE} outside the subsets.

\subsubsection{\ac{BIER-TE}-in-\ac{BIER-TE}}
Native \ac{TE} for subset tunneling can be achieved via \ac{BIER-TE}-in-\ac{BIER-TE} encapsulation.
Here, the \acp{S-BFIR} serve as \acp{BFER} for the tunneling \ac{BIER-TE} header.
When a \ac{S-BFIR} receives a packet, it removes the outer \ac{BIER-TE} header and forwards the packet based on the inner header.
However, this approach re-introduces the trade-off described in \sect{problem_statement}.
As illustrated in \fig{bierte_subset_scalability.pdf} the tunnel distribution trees to the subsets need to be small enough but still allow \ac{FRR} and \ac{TE}. 

\subsubsection{\ac{SR}/\ac{MPLS}}
Last, \ac{SR}/\ac{MPLS} tunnels can be used to implement the concept.
Similar to the other approaches, the original packet is encapsulated with \ac{SR}/\ac{MPLS}.
Then, it is forwarded to the \ac{S-BFIR} using the \ac{SR}/\ac{MPLS} forwarding logic.
When it reaches the \ac{S-BFIR} the header is removed and normal BIER-TE forwarding resumes.
In contrast to the other two approaches, it allows full \ac{TE} without any scalability issues.
Further, \ac{FRR} can be supported by using \ac{RSVP-TE}~\cite{rfc4090} or \ac{TI-LFA}~\cite{tilfa}.
However, \ac{SR}/\ac{MPLS} has to be supported on the forwarding nodes in the domain.

\par\medskip

We propose \ac{SR}/\ac{MPLS} tunnels as the preferred approach.
They support full \ac{TE} and \ac{FRR} on the path to the domain without scalability issues.
Further, these are common protocols supported in most networks.

%TODO: MPLS Egress Protection Framework! https://datatracker.ietf.org/doc/rfc8679/

\subsection{Ingress Protection}\label{sec:ingress_protection}

MPLS egress protection framework
Packet gets rerouted via tunnel, coinext ID determines oriogina ingress
Cannot use penultimate hop popping, because S-BFIR has to know that it is the secondary ingress

When an \ac{S-BFIR} fails, packets cannot be delivered to the \acp{BFER} of the corresponding subset anymore.
We therefore propose a subset ingress protection that leverages the existing \ac{FRR} mechanisms.
To that end, each subset must have more than one \ac{S-BFIR}.
Each \ac{S-BFIR} can be configured to serve as a backup ingress for one or more other \acp{S-BFIR}.

When an \ac{S-BFIR} fails, the preceding tunneling node becomes the \ac{PLR}.
It then reroutes the packet to one of the configured backup ingress \acp{S-BFIR}.
If \ac{MPLS} is used, this can be configured using \ac{RSVP-TE}~\cite{rfc4090} or \ac{TI-LFA}~\cite{tilfa}.

The backup \ac{S-BFIR} can detect rerouted packets by matching the links in the \ac{BS} that are connected to the protected \ac{S-BFIR}.
Each link of a protected \ac{S-BFIR} is added to the \ac{BTAFT} of the backup \ac{S-BFIR}.
These entries contain a backup path to the corresponding next hop.
If one of these entries matches, the backup ingress can forward the packet to the original next hops of the failed ingress.
Link bits of the backup \ac{S-BFIR} will not be matched for a rerouted packet since \acp{S-BFIR} are not transit nodes.

\subsection{Subset Selection}\label{sec:subset_selection}
An existing \ac{BIER-TE} network must be subdivided into subsets to use the proposed concept.
Therefore, an algorithm is required to decide which links and \acp{BFER} belong to which subsets.
Defining a concrete algorithm is out of scope for this work.
However, we present requirements for subsets that the algorithm has to adhere to: \ac{BS} length, 2-connectedness, backup ingresses and virtual links.

\subsubsection{\ac{BS} length}
The number of links and \acp{BFER} in a subset can not exceed the maximum supported length of the \ac{BS}.
This limit either stems from the hardware support or is defined by the network operator.

\subsubsection{2-connectedness}
If \ac{BIER-TE} \ac{FRR} is used, then each subset has to be at least two-connected.
Otherwise, there are no backup paths in the subset for link or node protection.
If the routing underlays \ac{FRR} is used, then the subsets do not have this requirement.

\subsubsection{Backup Ingresses}
For ingress protection, each subset has to have at least two \acp{S-BFIR}.
Further, these \acp{S-BFIR} should be located on the edge of a subset and not be used as transit nodes.

\subsubsection{Virtual Links}
For some topologies, ensuring that subsets are 2-connected is not possible.
As an example, a subset of nodes in a ring topology is only 2-connected when all nodes on the ring are in the same subset.
Therefore, virtual links over the routing underlay can be assigned between nodes of a subset.
These virtual links are then realized as \ac{BIER-TE} routed connections.

\par\medskip

In a given network, multiple subset layouts can fulfill the requirements.
Various optimization goals can be pursued, depending on the specific use case and network requirements.
One such goal could be to reduce the number of packet copies generated by the \ac{BFIR}.
To that end, destinations of the same \ac{IPMC} flow must be clustered in the same subset.
However, this requires knowledge about the expected flows in the network.
Another optimization goal could be the number of disjoint paths in a subset.
This allows a better load balancing within a subset.
\section{P4 Prototype Implementation}\label{sec:implementation}
%TODO: read implementation section
In this section, we introduce a prototype implementation for the proposed concept using P4 and the Intel Tofino\texttrademark~ASIC.
This prototype can act as any of the types of different forwarding nodes described in \sect{scalability}.
We first illustrate the implementation architecture and then elaborate on the processing logic for each supported protocol.
%We explain further additions made by this implementations

\subsection{Architecture}
The implementation is divided into four processing chains: BIER-TE, MPLS, IP, and port down.
When a packet enters the switch, it is passed to one of these chains based on the header information.
\fig{implementation.pdf} illustrates the implementation architecture and processing chains.

\figeps[\columnwidth]{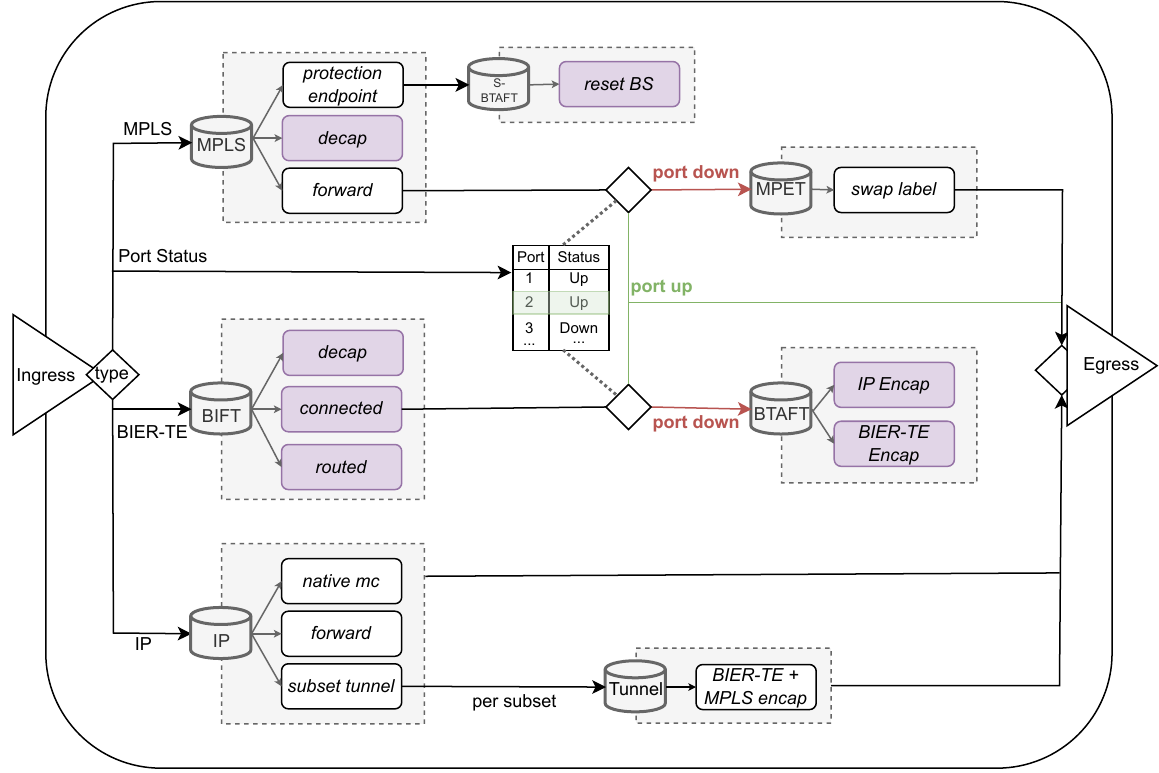}{The architecture of the \ac{BIER-TE} implementation, showing an abstract view of the P4 \acp{MAT}.
On ingress, a packet is assigned to one of the four possible processing chains based on header information.
The \acp{MAT} are visualized with their name and the available actions are listed next to them.
Actions that cause the packet to be recirculated are marked in purple with bold font.
}

The IP, MPLS, and BIER-TE chains implement the forwarding logic for the corresponding protocols.
A packet may be recirculated and pass through the same or another chain multiple times.
These protocol chains are discussed in more detail in sections \threesect{ipchain}{mplschain}{biertechain}.
The port down chain is responsible for tracking the state of each switch port in an internal register.
Each other chain can access this information to determine if the egress port of a packet is up or down.
The implementation is the same as presented in~\cite{MeLi2021} and will not be discussed further.

%\todo[inline]{Add egress processing section}
%\todo[inline]{Could precalculate all combinations of active NNH and resulting bitmasks.}

\subsection{IP Processing}\label{sec:ipchain}
In the IP chain, the initial \ac{MAT} matches the destination IP address of the packet.
If the packet matches, it is either forwarded as unicast, multicast, or encapsulated in the BIER-TE subset tunnel.
Processing subset tunnel encapsulation involves two steps and uses the egress processing of the Intel Tofino\texttrademark.
First, the original packet is copied for each destination subset using Tofino multicast groups.
Next, each packet is encapsulated with a distinct BIER-TE and MPLS tunnel header according to its destination subset during egress processing.

\subsection{MPLS Processing}\label{sec:mplschain}
In regular MPLS forwarding, the \ac{MAT} sets the egress port and sends the packet out of the switch.
Further, the \ac{MAT} also implements actions for MPLS tunnel endpoint handling.
If the packet has reached the subset tunnel endpoint, then the \ac{MAT} removes the MPLS label and recirculates the packet to the \ac{BIER-TE} chain.
If the packet has reached the MPLS egress protector, then the label is removed as well but the packet is passed to the egress processing.
Here, the \ac{S-BTAFT} matches on the bits of the next-hops of the failed node.
Its purpose is to apply the \ac{BIER-TE} FRR node protection, i.e., apply the \ac{BTAFT} processing that ensures that all next-hops of the failed node receive a packet copy.
It contains pre-computed entries with the reset and addBitmask for every possible combination of active next-hop bits.
This way, the packet does not have to be copied and recirculated for every active next-hop.

The \ac{MEPT} implements the MPLS egress protection and is only applied to the packet if there is a valid MPLS header.
If the egress port is down, then the label is swapped to the backup egress protection tunnel and transmitted over another port.
As of now, the implementation does not support label stacks.

\subsection{BIER-TE Processing}\label{sec:biertechain}
The structure of the \ac{BIFT} is as defined in the RFC~\cite{rfc9262} and implements the three forwarding actions connected, routed, and decap.
An incoming \ac{BIER-TE} packet has to be matched against the \ac{BIFT} for every active adjacency bit.
When a \ac{BIFT} entry matches on a packets BS, the entire packet is copied once.
The forwarding action is applied to the original packet and that packet is then passed to the \ac{BTAFT}.
The processed bits are removed from the BS in the copied packet and the packet is then recirculated.
This way, the succeeding \ac{BIFT} entries can match in the next iteration.
When no \ac{BIFT} entires match, all relevant adjacencies already received a packet copy, and the packet is discarded.

If a packet is forwarded via the connected forwarding function and the egress port is down, the \ac{BTAFT} is applied to that packet.
It matches on the egress port and active \ac{NNH} bits in the \ac{BS}.
For link protection, the packet will only match once and then be passed to egress.
For node protection, the \ac{BTAFT} contains an entry for every \ac{NNH}.
The packet is recirculated after a match for each active \ac{NNH} bit and the bit is unset in the original BS.
In each iteration, the implementation aggregates the add and resetBitmask of the matched entries.
When the packet does not match on any \ac{BTAFT} entry, all \ac{NNH} have been iterated.
Then, the aggregated bitmasks are applied to the original BS and the packet is recirculated to be handled by the \ac{BIFT}.

%After BIFT, BTAFT is applied as MAT if egress port is down
%Matches on next hop port (identifies failed node/link) and NNH bits in header
%If used for link protection, packet will match only once and then go to egress
%For node protection, table has entry for every NNH
%Therefore, has to match on the BTAFT once for each active NNH bit
%Implementation combines add and resetBitmask in each match on the table via bitwise AND of masks
%When all NNH are handled, aggregated masks are applied to original BS and packet is sent out to egress

%BIFT is built after RFC, the same functionality
%Problem has to match on BIFT multiple times, once for each active bit the switch is responsible for
%In P4 can only be achieved by recirculating a packet, is not possible to apply same table multiple times
%On each match same handling, BIFT action is applied to original packet and packet copy is recirculated
%Further, already handled bits are unset in the copy, otherwise infinite loop
%In each BIFT iteration, one of the three forwarding actions connected, routed or decap is applied to original packet
%When BIFT does not match, all adjacencies are handled, original packet is discarded

%\subsection{Discussion}

%Problem: Capacity required for recirculation reduces throughput
%E.g. if packet is recirculated three times, then 3x the traffic load
%Was already analyzed by lindner et al, show that can be improved using etc.

\section{Performance Evaluation}\label{sec:evaluation}
In this section, we evaluate the implementation of our P4 \ac{BIER-TE} prototype.
We first describe the evaluation setup and how the P4TG~\cite{p4tg} traffic generator P4TG is used.
% TODO: undo this change made for ht preprint Intel
%Then we measure the maximum throughput rate of ther P4 BIER-TE prototype in three scenarios: as BFIR, PLR and S-BFIR.
Then we measure the maximum throughput rate of ther P4 BIER-TE prototype as BFIR.
Further, we analyze the impact of the size of the \ac{BIER-TE} header and IPMC packet size on the implementations throughput.
Last, we summarize and discuss the evaluation.
%\todo[inline]{Anpassen für gekürzte Version}

\subsection{Evaluation Setup}
Our evaluation setup leverages the P4TG~\cite{p4tg} traffic generator that is connected to the implementation prototype.
P4TG can generate up to 100 Gbit/s of traffic per port and supports various protocols such as IP or MPLS.
It also provides extensive statistics for both generated and incoming traffic.
Since the original release, it has been further developed and ported to the Intel Tofino\texttrademark{} 2 that can generate up to 400 Gbit/s per port.
\fig{evaluation_setup_1.png} illustrates the evaluation setup.

%TODO: change to PDF!
\figeps[\columnwidth]{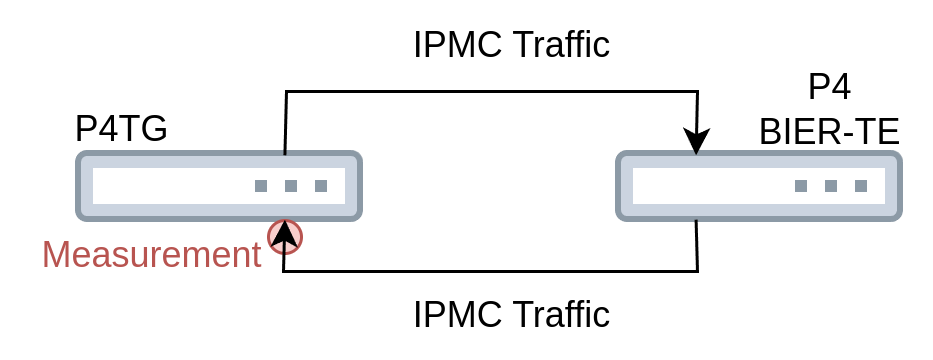}{The evaluation setup used in the following sections. The traffic generator P4TG is connected to the P4 BIER-TE prototype over a 100 Gbit/s link. Both applications are running on an Intel Tofino\texttrademark. P4TG sends IPMC traffic to the BIER-TE prototype and measures the returning traffic.}

We use P4TG to generate \ac{IPMC} traffic and measure the output of the P4 BIER-TE prototype.
Both P4TG and the BIER-TE prototype are running on separate Intel Tofino\texttrademark{} switching ASICs that are connected over a full duplex 100 Gbit/s link.
P4TG is configured to transmit 100 Gbit/s of constant bit-rate IPMC traffic to the BIER-TE prototype in each evaluation.
As P4TG does not support \ac{BIER} or \ac{BIER-TE} traffic, the experiments are set up such that the BIER-TE prototype always returns \ac{IPMC} traffic.
For each evaluation, we collect the maximum throughput measured by P4TG.

\subsection{BFIR Throughput}\label{sec:bfir}
We first evaluate the performance of the P4 prototype in the role of a \ac{BFIR}.
We compare the possible IPMC throughput for regular BIER-TE encapsulation and subset tunneling.
To that end, the P4 prototype is configured to add the \ac{BIER-TE} and MPLS headers to incoming IPMC traffic and recirculate it via a physical port\footnote{Standard recirculation causes the packet to enter the same port again and reduces the ingress capacity. Ports that are configured for physical recirculation do not share capacity with the original ingress port.}.
After the recirculation, the additional headers are removed and IPMC traffic is forwarded back to P4TG for rate measurement.
We first provide a formula to calculate the upper IPMC throughput limit for a given input frame size and BIER-TE \ac{BS} length.
Then, we show the collected measurements and analyze the results.

\subsubsection{Maximum Possible IPMC Throughput}\label{sec:throughput_formula}
A P4 program on the Intel Tofino\texttrademark can process data at a maximum rate of 100 Gbit/s per port\footnote{Programs that compile for the Intel Tofino\texttrademark{} always run at 100 Gbit/s processing speed per port.}.
We therefore provide a formula to calculate the maximum ingress IPMC data rate such that the egress data rate is smaller or equal to 100 Gbit/s.
To that end, we define $f_{max}$ as the rate of packets per second, with which the egress bandwidth is exactly 100 Gbit/s.
Given the size of incoming IPMC frames as $L_{IPMC}$ and the size of the BIER-TE header as $L_{BIER-TE}$ we can calculate the maximum input rate of IPMC traffic without packet loss $R_{max}$ as
\begin{equation}
    R_{max} = f_{max} * L_{IPMC} = \frac{100 \text{ } Gbit/s * L_{IPMC}}{L_{IPMC}+L_{BIER-TE}}
\end{equation}

\subsubsection{BFIR Throughput Measurements}
\twofigs{bierte_throughput.pdf}{bierte_subset_throughput.pdf} show the measured and calculated maximum IPMC throughput for a BIER-TE and subset tunneling BFIR.
In both figures, the traffic throughput increases with larger IPMC frame sizes since the packet rate reduces and thus the share of BIER-TE headers.
For a frame size of 1536 bytes, the throughput is above 99 Gbit/s for a BS length of 64 bit.
With the longest BS of 256 bits, the throughput decreases to around 96 Gbit/s.
If the incoming IPMC frames are as small as 64 bytes, the throughput decreases to 88 Gbit/s for a 64 bit BS and 70 Gbit/s for a 256 BS.

\twosubfigeps
{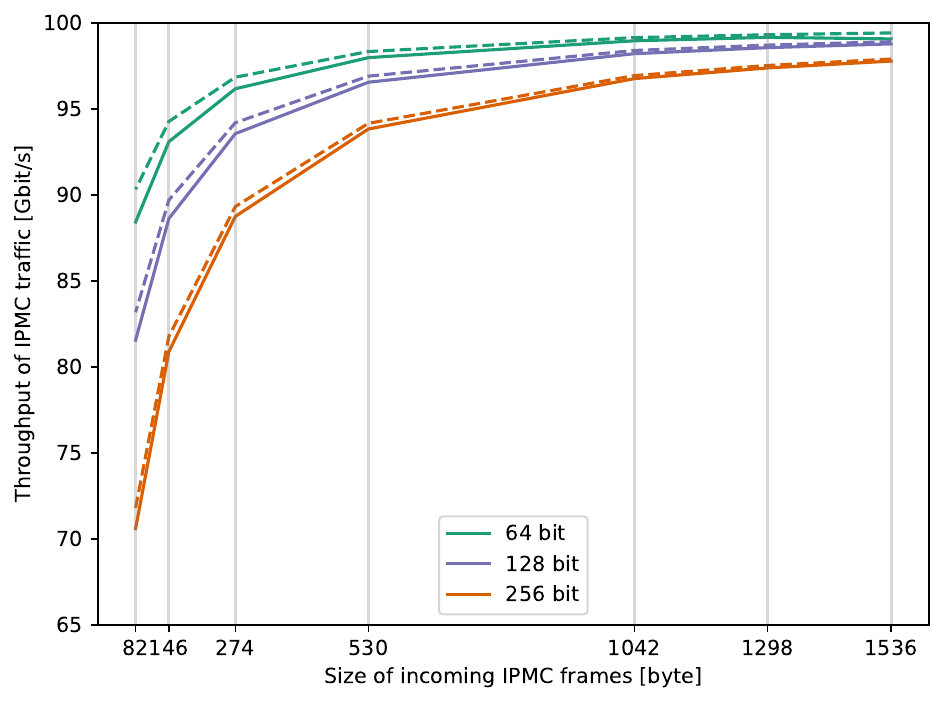}{BIER-TE BFIR.}
{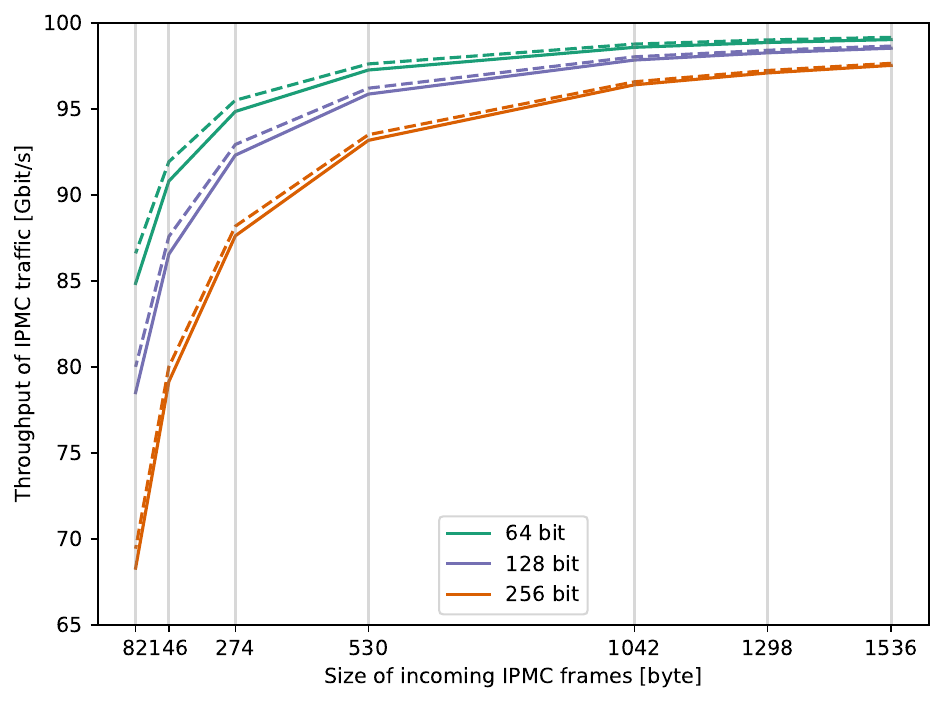}{subset tunneling BFIR.}
{Throughput rate of IPMC traffic for different BIER-TE BS lengths. The x-axis shows the size of incoming IPMC frames while the y-axis measures the throughput rate. The dashed lines represent the theoretical maximum throughput.}
%TODO: larger BS causes less difference between maximum and measurement

The figures show that for large frames, the implementation is around 0.3 Gbit/s below the theoretical maximum IPMC throughput.
This illustrates, that the P4 implementation gets very close to the maximum possible IPMC throughput and reaches close to 100 Gbit/s of BIER-TE traffic output.
Further, the figures demonstrate that the size of incoming packets has a strong impact on the possible IPMC throughput rate and may decrease it by up to 30\%.

\section{Conclusion}\label{sec:conclusion}
BIER-TE introduces tree engineering for BIER, a stateless transport method for IP multicast packets through multicast domains.
So far, BIER-TE could be applied only to small networks.
In this work, we presented architectural extensions for BIER-TE so that it can also be applied to large networks supporting resilient communications. 
For that purpose, we suggested subset tunneling. 
Ideally, it requires a division of a BIER domain into 2-connected subsets of receivers and links, possibly amended by virtual links. 
Ingress nodes of the BIER-TE domain tunnel BIER packets to subset ingresses from where packets are further delivered using normal BIER-TE forwarding. 
To meet resilience requirements, egress protection is utilized to protect against failures of subset ingresses, i.e., a point of local repair deviates a tunneled BIER packet to backup subset ingress. 
That node adopts BIER-TE-FRR in a modified way to act as a substitute for the failed subset ingress. 
The solution relies on well-known networking features where possible and requires only a few additions to existing standards. 
We implemented this architecture on the Intel Tofino ASIC and demonstrated that all components can forward traffic at a speed close to 100 Gb/s. 
Only replication nodes suffer from recirculations, which also holds for BIER, and we explain how this shortcoming can be mitigated by improvements we previously demonstrated for BIER. 
Future work should provide selection algorithms for optimal BIER-TE subsets including virtual links so that the proposed architectural extensions can be leveraged efficiently.

%\section{Reuse of BIER dataplane for BIER-TE}

%Standard BIER-TE: easier
%Bitmask meaning changes
%BFERs are still in bitmask, other bits are links
%Forwarding is still based on bit matching and applying FBM
%One small difference: Links have to be unset in bitmask before sneding over links

%Reuse in simple Scalability: easier
%Header meaning changes based on domain,
%Dataplane has to support Tunneling via IPv4 or BIER-te
%Domain Ingress is BFER in distribution (Backbone) domain
%After decapsulation, resubmission to forward based on inner Header

%Reuse in Scalability and FRR: harder
%FRR in distribution domain and subdomains same as normal FRR
%Problem: FRR for subdomain ingress BFRs
%If subdomain ingress fails, backup ingress has to receive packet

\begin{acronym}
% \acro{AVB}{Audio Video Bridging}
    \acro{BIER}{Bit Index Explicit Replication}
    \acro{BIER-TE}{Tree Engineering for Bit Index Explicit Replication}
    \acro{IPMC}{IP Multicast}
    \acro{FRR}{Fast Re-Route}
    \acro{TE}{Tree Engineering}
    \acro{BFIR}{Bit-Forwarding Ingress Router}
    \acro{BFER}{Bit-Forwarding Egress Router}
    \acro{BFR}{Bit-Forwarding Router}
    \acro{BIFT}{Bit Index Forwarding Table}
    \acro{BS}{BitString}
    \acro{F-BM}{Forwarding Bitmask}
    \acro{ECMP}{Equal Cost Multipath}
    \acro{P4}{programming protocol-independent packet processors}
    \acro{MAT}{match+action table}
    \acro{TNA}{Tofino native architecture}
    \acro{BTAFT}{BIER-TE Adjacency FRR Table}
    \acro{PLR}{Point of Local Repair}
    \acro{IGP}{Interior Gateway Protocol}
    \acro{SI}{Set Identifier}
    \acro{S-BF(IR}{Subdomain BFIR}
    \acro{SR}{Segment Routing}
    \acro{MPLS}{Multiprotocol Label Switching}
    \acro{RSVP-TE}{Resource Reservation Protocol - Traffic Engineering}
    \acro{LFA}{Loop Free Alternative}
    \acro{TI-LFA}{Topology Independant LFA}
    \acro{MRT}{Maximally Redundant Trees}
    \acro{MoFRR}{Multicast only FRR}
    \acro{S-BFIR}{Subset BFIR}
    \acro{MSR6}{Multicast Source Routing over IPv6}
    \acro{RBS}{Recursive BitString Structure}
    \acro{IETF}{Internet Engineering Task Force}
    \acro{MEPT}{MPLS Egress Protection Table}
    \acro{S-BTAFT}{Subset BTAFT}
    \acro{NNH}{Next-next Hop}
    \acro{SID}{Subset ID}
\end{acronym} 

\bibliography{conferences, literature} 
\bibliographystyle{ieeetr}

%\bibliographystyle{ieee} 
%\bibliography{bib/Literature}

\end{document}